\documentclass[twocolumn,aps,showpacs]{revtex4b4}
\usepackage{graphicx}
\begin{document}
\draft

\title {In-plane magnetic field-induced spin polarization and transition to insulating behavior
 in two-dimensional hole systems}

\author {E. Tutuc, E.P. De Poortere, S.J. Papadakis, and M. Shayegan}
\address{Department of Electrical Engineering, Princeton
University, Princeton, NJ 08544}
\begin{abstract}
Using a novel technique, we make quantitative measurements of the
spin polarization of dilute (3.4 to 6.8 $\times10^{10}$cm$^{-2}$)
GaAs (311)A two-dimensional holes as a function of an in-plane
magnetic field. As the field is increased the system gradually
becomes spin polarized, with the degree of spin polarization
depending on the orientation of the field relative to the crystal
axes. Moreover, the behavior of the system turns from metallic to
insulating \textit{before} it is fully spin polarized. The
minority-spin population at the transition is
$\sim8\times10^{9}$cm$^{-2}$, close to the density below which the
system makes a transition to an insulating state in the absence of
a magnetic field.
\end{abstract}
\maketitle

 A fundamentally important property of a two-dimensional (2D) carrier system is its
 response to a parallel magnetic field.
 Thanks to the Zeeman coupling, the parallel field induces a partial polarization of
 the carriers'  spin and yields a system with two separate spin subbands.
 Although there have been no direct measurements of the spin
 subband populations so far, certain features of the parallel
 magnetoresistance have been interpreted as the onset of full spin
 polarization of the 2D system \cite{Okamoto99, Mertes99, yoon00, Stergios:2D,Vitkalov}.
 The response to a parallel field may also help us understand why a 2D
 system, long believed to have an insulating ground state
 \cite{gangoffour, bishop}, displays a metallic
 behavior at finite temperatures in a certain range of carrier densities at zero magnetic field \cite{Okamoto99,
 Mertes99, yoon00, Stergios:2D,Vitkalov,krav,MIpapers,Murzin,Stergios:Sci,yaish,Stergios:symetry}.
 The parallel field also causes a transition from metallic to
 insulating behavior, and it has been suggested that this
 transition may be related to the spin polarization of the 2D
 system \cite{Mertes99, Stergios:2D, Vitkalov}. But again, there have been no quantitative
 measurements assessing this connection.

 Here we report direct and quantitative measurements of the spin polarization of
 dilute GaAs 2D hole systems (2DHSs) as a function of a strong
 parallel field ($B_{\|}$). We slowly rotate the sample in a
 large, almost parallel field [Fig. 1(a)], and measure its
 resistance as a function of the perpendicular component of the
 field ($B_{\bot}$). The latter leads to Shubnikov-de Haas (SdH)
 oscillations in the resistance, which we Fourier analyze to find
 the population of the spin subband densities and therefore the
 spin polarization of the 2DHS.
 Our experiments allow a determination of the spin polarization in a parallel field
 and of the field, $B_{P}$, above which the 2DHS
 becomes fully spin polarized. We find that both the
 spin polarization and $B_{P}$ depend on
 the orientation of $B_{\|}$ relative to the crystal axes, thus
 confirming the anisotropy of the $g$-factor in this system \cite{Stergios:2D,Winkler:g}.
 We also measure the temperature dependence of the
 resistance vs $B_{\|}$ traces to determine the field ($B_{I}$) above which
 the system makes a transition from metallic to insulating
 behavior. Combining these results, we determine the
 population of the minority spin subband at the field $B_{I}$. We
 find that in the density range of our study,
 the behavior turns insulating whenever the minority spin
 subband population drops below a threshold
 ($\sim8\times10^{9}$cm$^{-2}$), which is approximately
 independent of the total 2D hole density or the field $B_{I}$.
 Interestingly, this threshold population is of the order of the
 density below which our 2DHS makes a transition to an insulating behavior at
 zero magnetic field.

 \begin{figure*}
 \centering
 \includegraphics[scale=0.65]{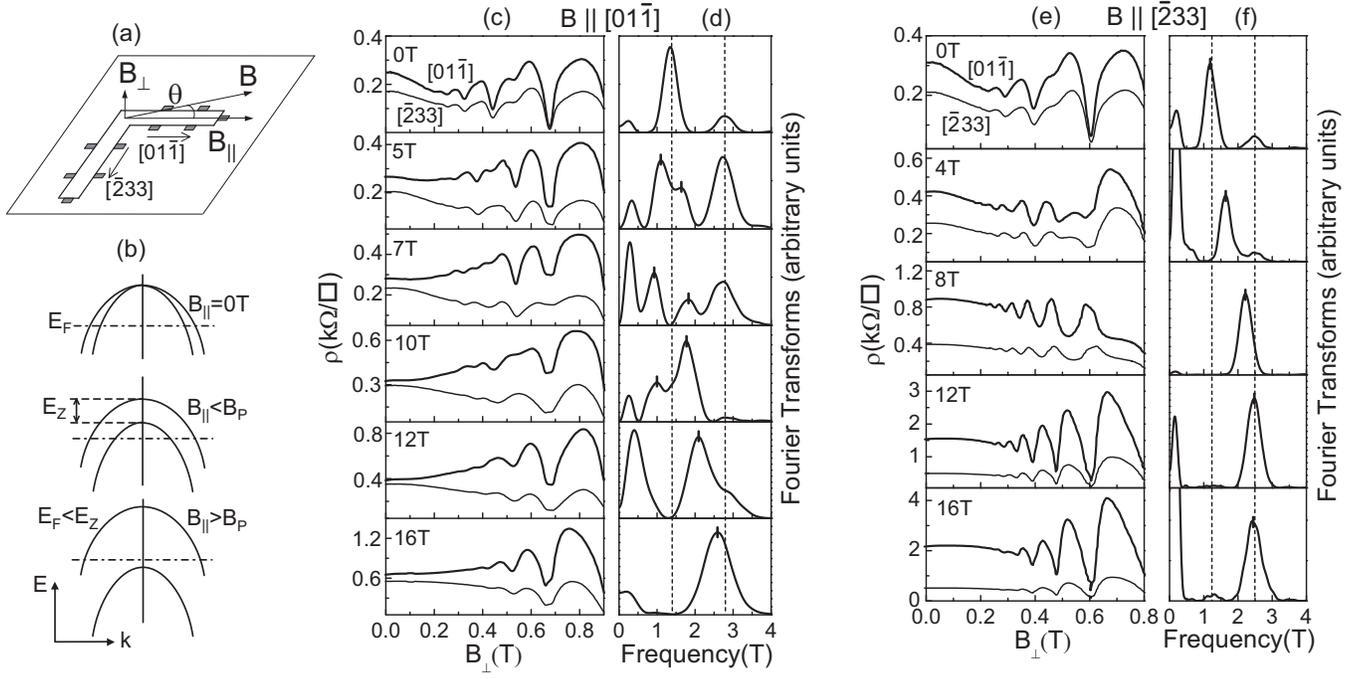}
 \caption {\small{(a) Experimental geometry. (b) Sketches of energy vs
 k-vector illustrating the evolution of spin subbands with
 increasing parallel field. (c) Resistivity vs $B_{\bot}$, for $p=6.8\times10^{10}$ cm$^{-2}$, at the indicated
 parallel fields applied along [$01\bar{1}$]. (d) Fourier transforms of the traces in (c).
 (e) and (f) Data similar to (c) and (d) but for $B$ $\parallel$ [$\bar{2}33]$ and $p=6.0\times10^{10}$ cm$^{-2}$.}}
 \label{raw-data}
 \end{figure*}

 We studied a Si-modulation doped GaAs quantum well
 grown on a (311)A GaAs substrate. The 2DHSs grown on GaAs (311)A
 substrates exhibit a mobility anisotropy stemming from an
 anisotropic surface morphology \cite{jean,wassermeier}. The
 interfaces between the GaAs and the AlGaAs are corrugated, with ridges
 along [$\bar{2}$33] direction, thus reducing the
 mobility for current parallel to [01$\bar{1}$]. We therefore used an
 L-shaped Hall bar aligned along [01$\bar{1}$] and [$\bar{2}$33]
 to allow simultaneous measurement of the resistivity along the two
 directions [Fig. 1(a)], and also deposited a metal front gate to control the
 density. We made measurements in pumped $^{3}$He at the temperature $T=0.3$K. The
 sample was mounted on a single-axis tilting stage that can be
 rotated, using a computer controlled stepper motor, in order to change the angle ($\theta$)
 between the sample plane and the magnetic field.
 The measurements were done using low-frequency
 lock-in techniques. As grown, the sample has a
 density  $p=6.8\times10^{10}$ cm$^{-2}$, and mobilities of 35 and 55 m$^{2}$/Vs along
 the [$01\bar{1}$] and [$\bar{2}33$] directions, respectively.

 Application of a magnetic field parallel to the 2DHS
 separates the two spin subbands by the Zeeman energy, $E_{Z}$, as shown schematically in Fig. 1(b) \cite{R2, R4}.
 In our experiments [Fig. 1(a)], we apply a constant magnetic
 field ($B$), slowly rotate the sample around $\theta=0^{\circ}$
 to introduce a small $B_{\bot}$, and measure both the
 longitudinal and Hall resistances. We use the Hall resistance
 to determine $B_{\bot}$. We then analyze the SdH oscillations
 induced by $B_{\bot}$, which give information about the Fermi surface of
 the 2DHS and the two spin subband densities. We emphasize that in our
 measurements, thanks to the high sample quality and the strength
 of the SdH oscillations, we are able to limit $B_{\bot}$ to
 sufficiently small values so that $B_{\|}\cong B$ to better than
 2.5\% during the sample rotation.

 In Fig. 1(c) we show plots of resistivity $\rho$ vs
 $B_{\bot}$, all taken at a fixed density. In each box, two traces are shown, one for sample
 current along [01$\bar{1}$] (thick line) and the other for [$\bar{2}$33]. The
 traces in the top box were taken with $\theta$ fixed at $90^{\circ}$
 and represent standard SdH measurements, i.e. $B_{\|}=0$T. From
 the corresponding Hall measurements (not shown), we determine a
 total hole density $p=6.8\times10^{10}$ cm$^{-2}$. The Fourier
 transform (FT) of the SdH oscillations for the [01$\bar{1}$]
 trace is shown in the top box Fig. 1(d); the FT of the
 [$\bar{2}$33] trace is similar and is not shown. The FT exhibits
 two peaks, one at 2.8T and another at half this
 value, at 1.4T \cite{R1}. The 2.8T peak, when multiplied by
 $(e/h)=2.42\times10^{10}$ cm$^{-2}/$T, gives $6.8\times10^{10}$
 cm$^{-2}$, equal to the total density of the 2DHS; this peak
 corresponds to the spin-resolved SdH oscillations, which are
 seen in the $\rho$ vs $B_{\bot}$ trace for
 $B_{\bot}>0.5$T. The main FT peak at 1.4T stems from the spin-unresolved oscillations.

 The traces in the rest of the boxes in
 Fig. 1(c) were taken at the indicated $B$ applied along [01$\bar{1}$] and by tilting
 the sample, as shown in Fig. 1(a). Note that in all
 traces, $B_{\|}\cong B$. With increasing $B_{\|}$, the
 lower (1.4T) peak in the FT spectrum splits into two peaks whose
 positions, when multiplied by $(e/h)$, add up to the total
 density of the sample. We therefore associate these two peaks
 with the Fermi contour areas (populations) \cite{R3} of the two Zeeman
 split spin subbands of the 2DHS [Fig. 1(b)]. As $B_{\|}$ increases,
 the majority spin subband peak merges with the total density (2.8T)
 peak, while the minority spin subband peak moves to very low
 frequencies and can no longer be resolved \cite{R1}.

 The panels (e) and (f) in Fig. 1 show similar data, but here $B_{\|}$ is
 along [$\bar{2}$33] and $p=6.0\times10^{10}$ cm$^{-2}$.
 The frequency of the FT peak associated with the majority spin subband density
 increases in value as $B_{\|}$ is increased from 4T to 12T and saturates at a
 constant value for $B_{\|}\geq12$T. This saturation signals the complete depletion of the minority
 spin subband and full spin polarization of the 2DHS.
 The saturation value of the FT peak position agrees with the
 total density of the 2DHS.

 \begin{figure}
 \includegraphics[scale=0.45]{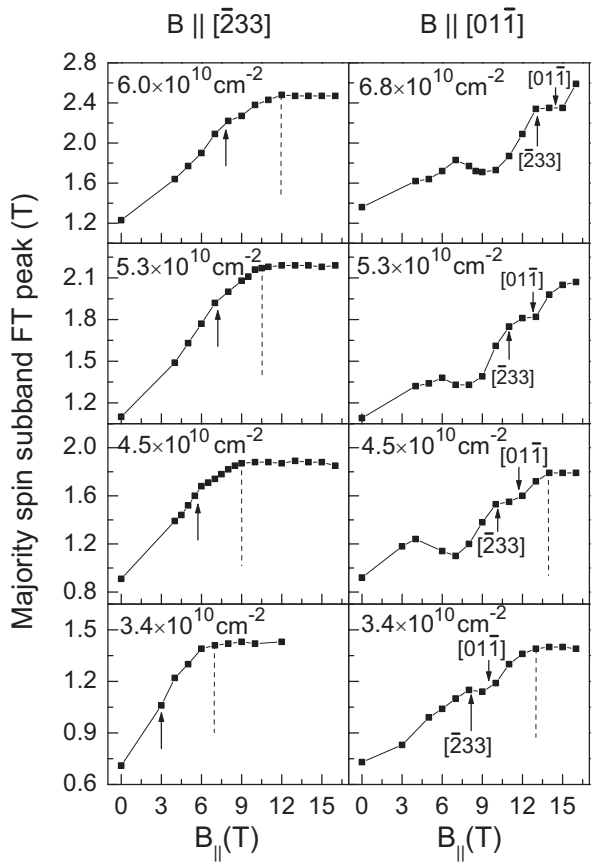}
 \caption{\small{The spin depopulation data for the two
 orientations of the parallel field at different densities.
 The dashed lines mark the field $B_{P}$ above which the 2D
 system is fully spin polarized.
 In each box, the arrows mark the field $B_{I}$ above which the 2D
 system turns insulating for the current along the indicated direction.
 In the left panels $B_{I}$ is independent of the current direction.}}
 \label{spin-results}
 \end{figure}

 In Fig. 2, we summarize the positions of the FT peaks corresponding
 to the majority spin subband as a function of $B_{\|}$ for the two cases examined in detail
 in Fig. 1 and  for three additional densities. There are several
 remarkable trends in these data. With increasing $B_{\|}$, the spin polarization of
 2DHS increases and saturates above a field $B_{P}$, marked
 by dashed vertical lines in Fig. 2. The field $B_{P}$ signals the
 full spin polarization of the 2D system and, as expected, $B_{P}$
 decreases with decreasing density \cite{R4}. However, $B_{P}$, as well as
 the rate at which the minority spin subband population decreases
 with $B_{\|}$, depend critically on the orientation of $B_{\|}$
 relative to the crystal axes. When $B_{\|}$ is applied along [$\bar{2}$33], the depopulation
 is monotonic and nearly linear with $B_{\|}$, but not so when $B_{\|}$ is along
 [01$\bar{1}$]. Moreover, for a given density, the depopulation is
 faster (smaller $B_{P}$) when $B_{\|}$ is parallel to
 [$\bar{2}$33]. These observations provide clear and quantitative
 evidence for the anisotropic dependence of the Zeeman energy and the $g$-factor on the orientation
 of parallel field in this system.

 The observed $g$-factor anisotropy is also qualitatively  consistent with the
 results of energy band calculations in this system \cite{Winkler:g}. The measured fields
 $B_{P}$, however, are about 2 to 3 times smaller than those
 calculated. We believe this discrepancy stems from the fact that
 many-body interactions are not included in the calculations. Such
 interactions are present in our dilute 2D system, and
 are expected to lead to an enhacement of the $g^{*}\cdot m^{*}$
 product \cite{Okamoto99, R4} and to lower the field $B_{P}$ at a given density. The
 non-monotonic $B_{\|}$-dependence of the spin polarization,
 for $B_{\|}$ along [01$\bar{1}$], is also quite intriguing and may be a result
 of many-body interactions: while band calculations exhibit some
 non-linearities in the spin polarization vs $B_{\|}$ curves
 \cite{Winkler:g, Winkler:unpublished}, they do not indicate the
 strong features observed in the measurements.

 \begin {figure}
 \includegraphics[scale=0.41]{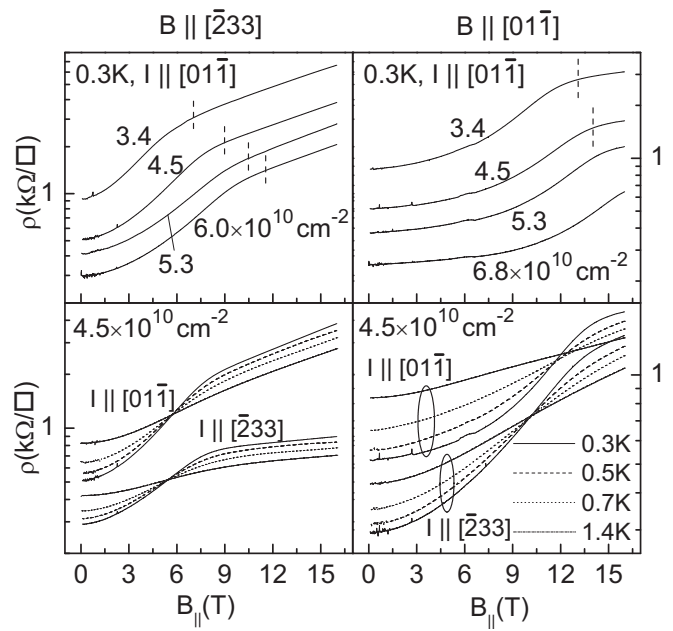}
 \caption {\small{Magnetoresistance vs $B_{\|}$ data at different indicated
 densities at 0.3K (top panels), and at a fixed density
 $4.5\times10^{10}$ cm$^{-2}$ and different temperatures (bottom
 panels). The dashed vertical lines in top panels mark the field $B_{P}$ above which the 2DHS
 becomes fully spin polarized.}}
 \label{MR-traces}
 \end {figure}

 We now present magnetoresistance (MR) data, taken as a function of
 $B_{\|}$ ($\theta=0^{\circ}$) and discuss their features in relation
 to the spin polarization data.  Such MR data have been recently
 reported for a number of 2D systems
 \cite{Okamoto99,Mertes99,yoon00,Stergios:2D,Vitkalov}. In Fig. 3 we provide
 examples of MR traces taken on the present sample, at different
 densities, for various orientations of $B_{\|}$ and current
 relative to the crystal axes.  In agreement with previous
 measurements on GaAs 2DHSs \cite{yoon00}, these traces
 exhibit a strong positive MR. The MR has a functional form
 $\sim e^{B^{2}/B_{2}^{2}}$ in the low field regime and $\sim e^{B/B_{1}}$ at high fields; the parameters
 $B_{1}$ and $B_{2}$ depend on density and orientation of $B_{\|}$ relative
 to the crystal axes. In Ref. \cite{Stergios:2D},
 the field around which the functional form changes is associated with the
 onset of full spin polarization.  Our measured $B_{P}$, marked with dashed
 vertical lines in Fig. 3 (top panels), are very close to the
 onset of the $e^{B/B_{1}}$ dependence and quantitatively validate
 the above association.

 Of particular relevance to the anomalous metallic-like behavior in
 2D systems is the $T$-dependence of the MR
 \cite{Mertes99,yoon00,Stergios:2D}.  In Fig. 3 (bottom
 panels) we show examples of such data for our sample at a density
 of $4.5\times10^{10}$ cm$^{-2}$.  The data, in agreement
 with previous results \cite{yoon00,Stergios:2D}, reveal that
 the metallic behavior observed at zero magnetic field becomes
 insulating above a certain field, $B_{I}$. For a given density,
 the field $B_{I}$ depends on the orientation of $B_{\|}$
 relative to both the crystal axes and the sample current
 direction [Figs. 2 and 3], but $B_{I}$ is always
 smaller than $B_{P}$, i.e., the transition to the insulating phase
 occurs before the 2DHS is fully spin polarized.  Our data
 allow us to answer an important question:  What is the population
 of the minority spin-subband at $B_{I}$, the onset of the
 insulating behavior?   In Fig. 4 we present a plot of the minority
 spin population at $B_{I}$ determined from our data.  This plot
 reveals a key result of our measurements:  the insulating behavior
 starts when the minority spin population drops below a threshold,
 $\sim 8\times10^{9}$ cm$^{-2}$, approximately independent of
 either the total density of the 2DHS, or the orientation of
 $B_{\|}$ relative to the crystal axes or the sample current
 direction! Also intriguing is the observation that, in the absence of an
 applied magnetic field, our 2DHS turns
 insulating when its total density falls below $\sim
 1\times10^{10}$ cm$^{-2}$.

 We conclude that the metallic behavior appears to be linked
 with the presence of two subbands whose populations are above a
 threshold. We suggest that a temperature dependent intersubband
 scattering mechanism may be responsible for the metallic
 behavior that we observe in this 2DHS at finite temperatures
 \cite{Murzin,Stergios:Sci,yaish,Stergios:2D}. When the density of one of these subbands
 falls below the $8\times10^{9}$ cm$^{-2}$
 threshold, its mobility is sufficiently reduced so that it stops
 to play a role in transport. This deduction is in fact
 corroborated by our own data
 as we see no sign of SdH oscillations of the minority spin-subband
 when the density of this subband is below the threshold. It is
 also interesting to note that the dependence of spin-polarization
 on $B_{\|}$ exhibits a kink near the field $B_{I}$ for both
 orientations of $B_{\|}$ relative to the crystal axes (Fig. 2).
 We hope that further experiments will shed light on this
 aspect of the data.

 \begin {figure}
 \includegraphics[scale=0.3]{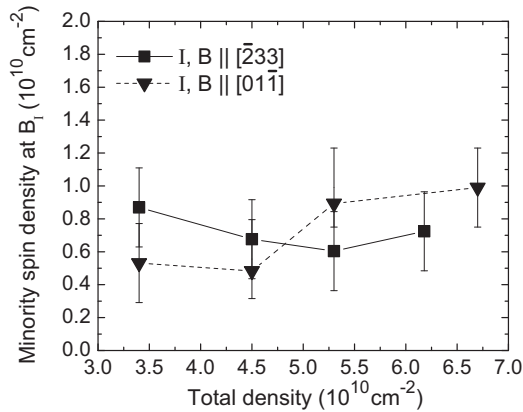}
 \caption {\small{The minority-spin subband density at the onset of insulating behavior plotted
 vs the total density of the 2DHS, for two different
 orientations of the parallel field and sample current. Data for the other current directions overlap
 these data within error bars.}}
 \label{summary}
 \end {figure}

 To summarize, via a novel technique, we have measured, the
 spin subband populations of a 2D hole system as a function of a parallel magnetic field.
 The system gradually becomes spin polarized, with the rate of
 polarization and the field of full spin polarization depending on the
 orientation of the parallel field relative to the crystal axes.
 Moreover, the 2D system turns insulating \textit{before} it
 is fully spin polarized, when its minority spin
 subband population drops below a threshold which is independent of the total density .

 Our work was supported by ARO, DOE, and NSF. We thank R. Winkler
 for valuable discussions.

 %\bibliography{spinbib}

\begin{thebibliography}{10}
\bibitem{R2}
 As a result of inversion asymmetry and strong spin-orbit coupling,
 the energy bands of GaAs 2D holes are spin-split even in the
 absence of a magnetic field (see, e.g., Ref. [11]). For
 completeness, we show this spin splitting in Fig. 1(b), even
 though the population difference between the two subbands is too
 small to be resolved in our samples, in the range of densities
 explored here.
\bibitem{R4}
 In a simplified picture, without spin-orbit coupling, the difference between the spin subbands populations is
 $\Delta p=(2\pi m^{*}/h^{2}) \cdot E_{Z}$, where
 $E_{Z}=\mu_{B}|g^{*}|B$ is the Zeeman energy, $m^{*}$ the
 effective mass, and $g^{*}$ the effective g-factor. In such a
 picture, the 2D system becomes fully spin polarized at a field
 $B_{P}=(h^{2}/2\pi\mu_{B})\cdot(p/m^{*}g^{*})$, where $p$ is the
 2D density.
\bibitem{R1}
 The peaks observed at very low frequencies ($<0.5$T) in Figs. 1(d) and 1(f) are artifacts.
\end{thebibliography}
 \end{document}